\documentclass[11pt]{article}
\usepackage[textwidth=15.2cm,textheight=22cm]{geometry}
\usepackage{amsmath,amssymb}
\usepackage{latexsym}
\usepackage{graphicx}
\usepackage{bbm}
\usepackage{cite}
\allowdisplaybreaks[1]
\numberwithin{equation}{section}

\newcommand{\be}{\begin{equation}}
\newcommand{\ee}{\end{equation}}
\newcommand{\ba}{\begin{eqnarray}}
\newcommand{\ea}{\end{eqnarray}}
\newcommand{\bdm}{\begin{displaymath}}
\newcommand{\edm}{\end{displaymath}}

\newcommand\fr[1]{\frac{1}{#1}}

\def\p{\partial}
\def\bp{\bar \partial}
\def\ba{\bar A}

\def\bh{\bar h}
\def\beq{\begin{equation}}
\def\eeq{\end{equation}}

\newcommand{\nn}{\nonumber}

\DeclareMathAlphabet{\mathpzc}{OT1}{pzc}{m}{it}

\newcommand{\ndt}{\noindent}

\def\bea{\begin{eqnarray}}
\def\eea{\end{eqnarray}}
\def\beas{\begin{eqnarray*}}
\def\eeas{\end{eqnarray*}}
\def\sla{\raise.15ex\hbox{$/$}\kern-.57em}

\def\parm{{\partial}_{-}}
\def\parp{\partial_+}

\def\spa#1.#2{\left\langle#1\,#2\right\rangle}
\def\spb#1.#2{\left[#1\,#2\right]}

\begin{document}

\begin{titlepage}
\vskip 1cm
\centerline{\LARGE {\bf {The quintic interaction vertex in light-cone gravity}}}

\vskip 2cm

\centerline{Sudarshan Ananth} 
\vskip .5cm
\centerline{\em  Indian Institute of Science Education and Research}
\centerline{\em Pune 411021, India}
\vskip 1.5cm

\vskip 1cm

\centerline{\bf {Abstract}}

\vskip .5cm

\noindent We consider pure gravity in light-cone gauge and derive the complete quintic interaction vertex. Up to quartic order, the Kawai-Lewellen-Tye (KLT) relations can be made manifest at the level of the Einstein-Hilbert Lagrangian. The quintic interaction vertex represents an essential first step in further extending the off-shell validity of the KLT relations to higher order vertices.

\end{titlepage}

\section{Introduction}

There are striking differences between the Lagrangians governing gravity and Yang-Mills theory. While the pure Yang-Mills Lagrangian involves only up to quartic interaction vertices, the gravity Lagrangian contains infinitely many interaction vertices. It is therefore rather surprising that there exist close perturbative ties between the two theories. In many ways, gravity seems to behave as the square of Yang-Mills
\beas
{\mbox {gravity}}\,\sim\,(\,{\mbox {Yang-Mills}}\,)\,\times\,(\,{\mbox {Yang-Mills}}\,)\ .
\eeas
One of the aims of this program of research is to make this statement completely precise. The Kawai-Lewellen-Tye (KLT) relations relate tree-level amplitudes in closed and open string theories~\cite{KLT}. In the field theory limit they relate scattering amplitudes in pure gravity to the ``square" of Yang-Mills amplitudes. For instance, in the case of three- and four-point amplitudes the KLT relations read
\bea
\label{kltrelate}
M^{\rm tree}_3(1,2,3)&&\!\!\!\!\!\!\!\!\!\!=A^{\rm tree}_3(1,2,3)\,A^{\rm tree}_3(1,2,3)\ , \nonumber \\
M^{\rm tree}_4(1,2,3,4)&&\!\!\!\!\!\!\!\!\!\!=-i\,s_{12}\,A^{\rm tree}_4(1,2,3,4)\,A^{\rm tree}_4(1,2,4,3)\ ,
\eea
where the $M_n$ represent gravity amplitudes and the $A_n$ are color-ordered amplitudes in pure Yang-Mills theory ($s_{ij}\equiv-(p_i+p_j)^2$). Tree-level amplitudes take a very compact form in a helicity basis so it is useful to work in light-cone gauge where only the helicity states propagate. Tree-level amplitudes in which precisely two external legs carry negative helicity are called maximally helicity violating (MHV) amplitudes~\cite{LD}. 
\vskip 0.2cm
\ndt The Einstein-Hilbert Lagrangian does not exhibit any obvious factorization property that might explain the origin of the perturbative relations in (\ref {kltrelate}). This raises two interesting questions
\vskip 0.2cm
\ndt a. Can we derive these KLT relations directly from the Einstein-Hilbert Lagrangian?$\\$
b. Do the KLT relations hold only for on-shell scattering amplitudes or, more generally, at
\vskip 0.05cm
\ndt $\;\;\;\;\;$the level of the Lagrangian~\footnote{If the KLT relations are indeed valid at the level of the Lagrangian (valid off-shell) to all orders, then they can be used directly in loop computations which have so far required unitarity-based techniques~\cite{BDR}.}?
\vskip 0.2cm
\ndt In~\cite{AT}, the three- and four-point KLT relations were made manifest at the level of the Lagrangian thus providing a partial answer to both questions (see also~\cite{related}). However, all non-vanishing amplitudes at the three- and four-point levels are purely MHV making the results of~\cite{AT} MHV-specific. It is therefore important to examine higher order vertices where non-MHV structures appear. In this paper we derive the complete quintic interaction vertex in light-cone gravity. This represents a necessary first step in the process of extending the analysis of~\cite{AT} to the next order~\footnote{Note that the truly non-MHV structures first appear only in the six-point vertex, a point we will return to later in the paper.}.
\vskip 0.2cm
\ndt The quintic interaction vertex in light-cone gravity is an interesting result, by itself, for two other reasons: ({\bf {a}}) recent results seem to suggest that pure gravity is better behaved in the ultra-violet than previously believed~\cite{BCFIJ}. This could signal the presence of a hidden symmetry that leads to cancellations beyond those expected from power counting. It will be fascinating to look for signs of such a symmetry by combining loop calculations in light-cone gauge~\cite{MO} with group-theoretical techniques~\cite{TC}. ({\bf {b}}) the five-point interaction vertex is crucial to determining the corresponding interaction vertex in light-cone superspace for $\mathcal N=8$ supergravity (the quartic interaction vertex of pure gravity was used in~\cite{ABHS} to obtain the
corresponding vertex for $\mathcal N=8$ supergravity in light-cone superspace). A complete light-cone superspace formulation will allow us to apply the techniques of~\cite{SM,LB} to $\mathcal N=8$ supergravity~\footnote{These techniques, primarily based on power counting, are not very useful for a non-renormalizable theory, however, such an analysis may offer additional insight into the results of~\cite{Bern}}.

\vskip 0.5cm
\section{Gravity in light-cone gauge}

We begin with a brief review of the light-cone formulation of pure gravity. This will lead us to a closed form expression for the Lagrangian. From this closed form, we will extract the five-graviton interaction vertices.

\vskip 0.2cm

\ndt The derivation of the closed form Lagrangian is based on the formalism developed in~\cite{SS}. A perturbative expansion of this Lagrangian, to order $\kappa^2$, was performed in~\cite{BCL}. This section offers a quick review of those results.

\vskip 0.2cm
\ndt With the metric $(-,+,+,+)$ we define
\be
x^\pm\,=\,\fr{\sqrt 2}\,(x^0\,\pm\,x^3)\ , \quad \partial_\pm\,=\,\fr{\sqrt 2}\,(\partial_0\,\pm\,\partial_3)\ .
\ee
$x^+$ plays the role of light-cone time and $-i\,\partial_+$ the light-cone Hamiltonian. $\parm$ is now a spatial derivative and its inverse, $\frac{1}{\parm}$, is defined using the prescription in~\cite{SM}. The transverse coordinates and their derivatives read
\bea
x&&\!\!\!\!\!\!\!\!=\fr{\sqrt 2}\,(x^1\,+i\,x^2)\ , \quad {\bar \partial}\equiv\frac{\partial}{\partial x}=\fr{\sqrt 2}\,(\partial_1\,-\,i\,\partial_2)\ , \nonumber \\
{\bar x}&&\!\!\!\!\!\!\!\!=\fr{\sqrt 2}\,(x^1\,-i\,x^2)\ , \quad \partial\equiv\frac{\partial}{\partial {\bar x}}=\fr{\sqrt 2}\,(\partial_1\,+\,i\,\partial_2)\ .
\eea

\ndt The Einstein-Hilbert action is
\bea
S_{EH}=\int\,{d^4}x\,{\cal L}\,=\,\frac{1}{2\,\kappa^2}\,\int\,{d^4}x\,{\sqrt {-g}}\,R\ ,
\eea
where $g=\det{g_{\mu\nu}}$, $R$ is the curvature scalar and the covariant equations of motion are $R_{\mu\nu}=0$.
\vskip 0.2cm
\ndt Light-cone gauge is chosen by setting
\bea
g_{--}\,=\,g_{-i}\,=\,0\ ,\qquad \mbox {\footnotesize $i=1,2$}\ .
\eea
with constraint relations following from $R_{-i}=0$ and $R_{--}=0$~\cite{SS}. The Lagrangian density now reads
\bea
\label{lclagden}
{\cal L}=\fr{2\kappa^2}\,\sqrt{-g}\;\bigg( 2 g^{+-} R_{+-} +g^{ij} R_{ij}\bigg)\ .
\eea
The metric is parameterized as follows
\bea
\label{param}
g_{+-}\,=\,-\,e^\frac{\psi}{2}\ ,\quad g_{ij}\,=\,e^\psi\,\gamma_{ij}\ .
\eea
The field $\psi$ is real while $\gamma_{ij}$ is a $2\times 2$ real, symmetric, unimodular matrix. $R_{-i}=0$ determines $g^{-i}$ while $R_{--}=0$ gives 
\bea
\label{psi}
\psi=\fr{4}\,\fr{\parm^2}\,\big(\parm \gamma^{i j} \parm \gamma_{ij}\big)\ .
\eea
In terms of (\ref {param}), the Lagrangian density in (\ref {lclagden}) is~\cite{SS,BCL}
\bea
\label{finallag}
{\cal L}&=&\fr{2\kappa^2}\bigg\{\;e^{\psi}\bigg(\frac{3}{2}\parp\parm\psi-\fr{2}\parp\gamma^{ij}\parm\gamma_{ij}\bigg) \nonumber \\
&&-e^{\frac{\psi}{2}}\gamma^{ij}\bigg(\fr{2}\p_i\p_j\psi-\frac{3}{8}\p_i\psi\p_j\psi-\fr{4}\p_i\gamma^{kl}\p_j\gamma_{kl}+\fr{2}\p_i\gamma^{kl}\p_k\gamma_{jl}\bigg) \nonumber \\
&&-\fr{2}e^{-\frac{3}{2}\psi}\gamma^{ij}\fr{\parm}R_i\fr{\parm}R_j\,\Bigg\}\ ,
\eea
where
\bea
R_i=e^{\psi}\bigg(-\fr{2}\parm\gamma^{jk}\p_i\gamma_{jk}+\frac{3}{2}\parm\p_i\psi-\fr{2}\p_i\psi\parm\psi\bigg)-\p_k\bigg(e^{\psi}\gamma^{jk}\parm\gamma_{ij}\bigg)\ .
\eea
This is the closed form of the light-cone Lagrangian for pure gravity. Given (\ref {psi}), we see that the Lagrangian is a function of exactly two fields (since $\gamma_{ij}$ is unimodular and symmetric) which correspond to the physical degrees of freedom of the graviton.

\vskip 0.5cm
\subsection{Perturbative expansion}

A perturbative expansion of the Lagrangian in (\ref {finallag}) was obtained in~\cite{BCL} by choosing
\bea
\label{gamma}
\gamma_{ij}=\left(e^{\kappa h}\right)_{ij}\ ,\qquad h_{ij}=\fr{\sqrt 2}\begin{pmatrix} h_{11} & h_{12}\\h_{12} &-h_{11}\end{pmatrix}\ .
\eea
The idea here is to expand (\ref {finallag}) in powers of $\kappa$ thereby obtaining an expansion in terms of vertices which involve an increasing number of fields. We identify the physical helicity states of the graviton by
\bea
\label{grav}
h_{ij}=\fr{\sqrt 2}\begin{pmatrix} h+\bh & -i(h-\bh)\\-i(h-\bh) &-h-\bh\end{pmatrix}\ ,
\eea
with $h$ and $\bh$ representing gravitons of helicity $+2$ and $-2$ respectively.
\vskip 0.2cm
\ndt The light-cone Lagrangian density for pure gravity, to order $\kappa^2$ is~\cite{ABHS,SS,BCL}
\bea
\label{lcg}
{\cal L}\;\;=&&\!\!\!\!\!\!\bh\, \square \, h+\kappa\,\bh\,\parm^2\bigg(\frac{\bp}{\parm}h\frac{\bp}{\parm}h-h\frac{\bp^2}{\parm^2}h\bigg)+\kappa\,h\,\parm^2\bigg(\frac{\p}{\parm}\bh\frac{\p}{\parm}\bh-\bh\frac{\p^2}{\parm^2}\bh\bigg) \nonumber \\
&&\!\!\!\!\!\!\!\!\, \nonumber \\
&&\!\!\!\!\!\!\!\!+\kappa^2{\biggl \{}\,\fr{\parm^2}\big(\parm h\parm\bh\big)\frac{\p\bp}{\parm^2}\big(\parm h\parm\bh\big)+\fr{\parm^3}\big(\parm h\parm\bh\big)\left(\p\bp h\,\parm\bh+\parm h\p\bp\bh\right) \nonumber\\
&&\!\!\!\!\!\!\!\!-\fr{\parm^2}\big(\parm h\parm\bh\big)\,\left(2\,\p\bp h\,\bh+2\,h\p\bp\bh+9\,\bp h\p\bh+\p h\bp\bh-\frac{\p\bp}{\parm}h\,\parm\bh-\parm h\frac{\p\bp}{\parm}\bh\right) \nonumber  \\
&&\!\!\!\!\!\!\!\!-2\fr{\parm}\big(2\bp h\,\parm\bh+h\parm\bp\bh-\parm\bp h\bh\big)\,h\,\p\bh-2\fr{\parm}\big(2\parm h\,\p\bh+\parm\p h\,\bh-h\parm\p\bh\big)\,\bp h\,\bh \nonumber \\
&&\!\!\!\!\!\!\!\!-\fr{\parm}\big(2\bp h\,\parm\bh+h\parm\bp\bh-\parm\bp h\bh\big)\fr{\parm}\big(2\parm h\,\p\bh+\parm\p h\,\bh-h\parm\p\bh\big) \nonumber \\
&&\!\!\!\!\!\!\!\!-h\,\bh\,\bigg(\p\bp h\,\bh+h\p\bp\bh+2\,\bp h\p\bh+3\frac{\p\bp}{\parm}h\,\parm\bh+3\parm h\frac{\p\bp}{\parm}\bh\bigg){\biggr \}}\ .
\eea
In obtaining this result, order $\kappa^2$ terms containing a $\parp$ have been removed using the following redefinition~\cite{ABHS}
\bea
\label{shift}
h\rightarrow h-\kappa^2\,\fr{\parm}{\biggl \{}2\,\parm^2h\fr{\parm^3}(\parm h\parm\bh)+\parm h\fr{\parm^2}(\parm h\parm\bh)+\fr{3}(h\bh\parm h-hh\parm\bh)\,{\biggr \}}\ .
\eea
This is the Lagrangian density to order $\kappa^2$ first derived in~\cite{BCL} and used in~\cite{AT}.

\vskip 0.5cm
\subsection{Simpler variables}

Having reviewed the light-cone description of gravity to order $\kappa^2$, we now focus on expanding (\ref {finallag}) to order $\kappa^3$. Before doing so, we simplify the notation a little by choosing to work directly with the transverse coordinates. We introduce indices $a,b$ such that $x^a$ represents $x,\bar x$. The corresponding metric is
\bea
\eta_{ab}=\frac{\p x^i}{\p x^a}\,\frac{\p x^j}{\p x^b}\,\eta_{ij}\ .
\eea
So 
\bea
\eta_{xx}=\eta_{\bar x\bar x}=0\ ,\qquad \eta_{x{\bar x}}=\eta_{{\bar x}x}=1\ .
\eea
The inverse metric $\eta^{ab}$ has components
\bea
\eta^{xx}=\eta^{\bar x\bar x}=0\ ,\qquad \eta^{x{\bar x}}=\eta^{{\bar x}x}=1\ .
\eea
The matrix representation of $h$ is now
\bea
h_{ab}={\sqrt 2}\,\begin{pmatrix} \bh & 0\\0 & h\end{pmatrix}\ .
\eea
Our aim is to retain, in the expansion of (\ref {finallag}), all terms of the form
\bea
h^p\,\bh^q\ ,\qquad p+q\,\leq\,5\ .
\eea
From the closed form structure in (\ref {finallag}), it is clear that we only need to compute the quantities $\psi$ and $\gamma$, from (\ref {psi}) and (\ref {gamma}), to order $p+q\,\leq\,4$ for this purpose~\footnote{$\psi$ being an even function does not contain a fifth order term}. To this order, the relevant expressions are
\bea
\label{matrixone}
\gamma_{ab}=\begin{pmatrix} {\sqrt 2}\,\bh+\frac{\sqrt 2}{3}\bh\bh h & 1+\bh h+\fr{6}\bh\bh hh\\\\1+\bh h+\fr{6}\bh\bh hh & {\sqrt 2}\,h+\frac{\sqrt 2}{3}\bh hh\end{pmatrix}\ ,
\eea
\bea
\label{matrixtwo}
\gamma^{ab}=\begin{pmatrix} -{\sqrt 2}\,h-\frac{\sqrt 2}{3}\,\bh hh & 1+\bh h+\fr{6}\bh\bh hh\\\\1+\bh h+\fr{6}\bh\bh hh & -{\sqrt 2}\,\bh-\frac{\sqrt 2}{3}\,\bh\bh h \end{pmatrix}\ ,
\eea
and
\bea
\label{theone}
\psi\!=\!-\fr{\parm^2}(\parm \bh\parm h)\!-\!\fr{3\parm^2}(\parm h\parm[\bh\bh h])\!-\!\fr{3\parm^2}(\parm[\bh hh]\parm\bh)\!+\!\fr{2\parm^2}(\parm[\bh h]\parm[\bh h])\ .
\eea
In the next section, we use these results to derive the entire quintic interaction vertex.

\vskip 0.5cm
\section{The quintic interaction vertex}

From (\ref {finallag}), we see that each term involves at most two transverse derivatives. Thus helicity considerations only permit two kinds of terms at fifth order: $h^3\bh^2$ and $h^2\bh^3$. 
\vskip 0.2cm
\ndt We start from the closed form Lagrangian in (\ref {finallag}) and extract all vertices that involve five graviton interactions. We illustrate the procedure with a specific example: consider the fifth term in (\ref {finallag})
\bea
\fr{4}e^\frac{\psi}{2}\gamma^{ij}\p_i\gamma^{kl}\p_j\gamma_{kl}\ .
\eea
Expansion of the $kl$ indices yields
\bea
\fr{4}e^\frac{\psi}{2}\gamma^{ij}\p_i\gamma^{xx}\p_j\gamma_{xx}+\fr{4}e^\frac{\psi}{2}\gamma^{ij}\p_i\gamma^{\bar x\bar x}\p_j\gamma_{\bar x\bar x}+\fr{2}e^\frac{\psi}{2}\gamma^{ij}\p_i\gamma^{x\bar x}\p_j\gamma_{x\bar x}\ .
\eea
Further expansion followed by the use of (\ref {matrixone}), (\ref {matrixtwo}) and (\ref {theone}) results in
\beas
&&\!\!\!\!\!\!\!\!\!\!\!-\fr{\sqrt 2}h\bp h\bp\bh\fr{\parm^2}(\parm\bh\parm h)\!+\!\frac{\sqrt 2}{3}\bh hh\bp h\bp\bh\!+\!\frac{\sqrt 2}{3}h\bp h\bp(\bh\bh h)\!+\!\frac{\sqrt 2}{3}h\bp\bh\bp(\bh hh)\!-\!\fr{\sqrt 2}h\bp(\bh h)\bp(\bh h) \\
&&\!\!\!\!\!\!\!\!\!\!\!-\fr{\sqrt 2}\bh\p h\p\bh\fr{\parm^2}(\parm\bh\parm h)\!+\!\frac{\sqrt 2}{3}\bh\bh h\p h\p\bh\!+\!\frac{\sqrt 2}{3}\bh\p h\p(\bh\bh h)\!+\!\frac{\sqrt 2}{3}\bh\p\bh\p(\bh hh)\!-\!\fr{\sqrt 2}h\p(\bh h)\p(\bh h)\ .
\eeas

\ndt Detailed calculations for each individual term in (\ref {finallag}) are not shown here. Instead, we simply present the final result below. The complete quintic interaction vertex for pure gravity in light-cone gauge is
\bea
{\cal L}_{\kappa^3}=\kappa^3\,A+\kappa^3\,{\bar A}\ ,
\eea
where the expression for $A$ reads ($\bar A$ is the complex conjugate of $A$)
\bea
\label{result}
A=&&\!\!\!\!\!\!-\fr{\sqrt 2}h\bp h\bp\bh\fr{\parm^2}(\parm\bh\parm h)+\frac{\sqrt 2}{3}\bh hh\bp h\bp\bh+\frac{\sqrt 2}{3}h\bp h\bp(\bh\bh h)+\frac{\sqrt 2}{3}h\bp\bh\bp(\bh hh) \nn \\
&&\!\!\!\!\!\!+\fr{\sqrt 2}h\bp h\bp\bh\fr{\parm^2}(\parm\bh\parm h)-\frac{\sqrt 2}{3}\bh hh\bp h\bp\bh-\frac{\sqrt 2}{3}h\bp h\bp(\bh\bh h)-\frac{\sqrt 2}{3}h\bp\bh\bp(\bh hh) \nn \\
&&\!\!\!\!\!\!-\frac{3}{4\sqrt 2}h\frac{\bp}{\parm^2}(\parm\bh\parm h)\frac{\bp}{\parm^2}(\parm\bh\parm h)+\fr{2\sqrt 2}h\fr{\parm^2}(\parm\bh\parm h)\frac{\bp\bp}{\parm^2}(\parm\bh\parm h) \nn \\
&&\!\!\!\!\!\!-\fr{3\sqrt 2}\bh hh\frac{\bp\bp}{\parm^2}(\parm\bh\parm h)-\fr{3\sqrt 2}h\frac{\bp\bp}{\parm^2}(\parm h\parm[\bh\bh h])-\fr{3\sqrt 2}h\frac{\bp\bp}{\parm^2}(\parm[\bh hh]\parm\bh)\nn \\
&&\!\!\!\!\!\!-\fr{2\sqrt 2}h\frac{\bp\bp}{\parm^2}(\parm[\bh h]\parm[\bh h])-2{\sqrt 2}\,\bh\bp h{\biggl [}\frac{\bp}{\parm}\{h\parm(\bh h)\}+\frac{\bp}{\parm}\{\fr{\parm^2}(\parm h\parm\bh)\parm h\} \nn \\
&&\!\!\!\!\!\!-\frac{\bp}{\parm}(h\bh\parm h)-\fr{3}\bp(hh\bh)+\fr{\parm}\{\fr{\parm^2}(\parm h\parm\bh)\bp\parm h\}\,{\biggr ]}+\fr{\sqrt 2}\,h\,\fr{\parm}{\biggl [}\parm h\bp\bh \nn \\
&&\!\!\!\!\!\!+\parm\bh\bp h-\frac{3}{2}\frac{\bp}{\parm}(\parm\bh\parm h)+2\bp(h\parm\bh)-\bp\parm(\bh h){\biggr ]}\,\times\,\fr{\parm}{\biggl [}\parm h\bp\bh+\parm\bh\bp h \nn \\
&&\!\!\!\!\!\!-\frac{3}{2}\frac{\bp}{\parm}(\parm\bh\parm h)+2\bp(h\parm\bh)-\bp\parm(\bh h){\biggr ]}+\bp h\bp h\,{\biggl [}\,\frac{\sqrt 2}{3}\bh\bh h+\frac{3}{\sqrt 2}\bh\fr{\parm^2}(\parm h\parm\bh)\,{\biggr ]} \nn \\
&&\!\!\!\!\!\!+{\sqrt 2}\bp h\,\fr{\parm}{\biggl \{}-\fr{\parm^2}(\parm h\parm\bh)\parm h\bp\bh+\fr{3}\parm h\bp(\bh\bh h)+\fr{3}\parm(\bh hh)\bp\bh+\fr{3}\parm\bh\bp(\bh hh) \nn \\
&&\!\!\!\!\!\!-\fr{\parm^2}(\parm h\parm\bh)\parm\bh\bp h+\fr{3}\parm(\bh\bh h)\bp h-\parm(\bh h)\bp(\bh h)-\fr{2}\frac{\bp}{\parm}[\parm h\parm(\bh\bh h)] \nn \\
&&\!\!\!\!\!\!+\frac{3}{2}\fr{\parm^2}(\parm h\parm\bh)\frac{\bp}{\parm}(\parm\bh\parm h)-\fr{2}\frac{\bp}{\parm}[\parm(\bh hh)\parm\bh]+\frac{3}{4}\frac{\bp}{\parm}[\parm(\bh h)\parm(\bh h)] \nn \\
&&\!\!\!\!\!\!-\fr{2}\frac{\bp}{\parm^2}(\parm\bh\parm h)\fr{\parm}(\parm\bh\parm h)+\frac{2}{3}\bp[h\parm(\bh\bh h)]+\frac{2}{3}\bp[\bh hh\parm\bh]-\fr{6}\bp\parm(\bh\bh hh) \nn \\
&&\!\!\!\!\!\!-2\bp[\fr{\parm^2}(\parm h\parm\bh)h\parm\bh]-\bp[\bh h\parm(\bh h)]+\bp[\fr{\parm^2}(\parm\bh\parm h)\parm(\bh h)]\,{\biggr \}} \nn \\
&&\!\!\!\!\!\!-{\sqrt 2}\fr{\parm}{\biggl \{}\fr{\parm^2}(\parm h\parm\bh)\bp\parm h-\fr{3}\bp\parm(hh\bh)-\bp(h\bh\parm h)+\bp[\fr{\parm^2}(\parm h\parm\bh)\parm h] \nn \\
&&\!\!\!\!\!\!+\bp(h\parm(\bh h){\biggr \}}\times\,\fr{\parm}{\biggl \{}\parm h\bp\bh+\parm\bh\bp h-\frac{3}{2}\frac{\bp}{\parm}(\parm\bh\parm h)+2\bp(h\parm\bh)-\bp\parm(\bh h)\,{\biggr \}} \nn \\
\, \nn \\
&&\!\!\!\!\!\!\!\!\!\!\!\!+{\sqrt 2}[\bh h+\frac{3}{2}\fr{\parm^2}(\parm h\parm\bh)]\,\fr{\parm}{\biggl \{}\parm h\bp\bh+\parm\bh\bp h-\frac{3}{2}\frac{\bp}{\parm}(\parm\bh\parm h)+2\bp(h\parm\bh)-\bp\parm(\bh h)\,{\biggr \}}\bp h \nn \\
&&\!\!\!\!\!\!\!\!\!\!\!\!+{\biggl \{}2\,\parm^2\bh\fr{\parm^3}(\parm h\parm\bh)+\parm\bh\fr{\parm^2}(\parm h\parm\bh)+\fr{3}(h\bh\parm \bh-\bh\bh\parm h)\,{\biggr \}}\parm\bigg(\frac{\bp}{\parm}h\frac{\bp}{\parm}h-h\frac{\bp^2}{\parm^2}h\bigg) \nn \\
&&\!\!\!\!\!\!\!\!\!\!\!\!+{\biggl \{}2\,\parm^2h\fr{\parm^3}(\parm h\parm\bh)+\parm h\fr{\parm^2}(\parm h\parm\bh)+\fr{3}(h\bh\parm h-hh\parm\bh)\,{\biggr \}}\times{\biggl [}2\frac{\bp}{\parm^2}(\parm^2\bh\frac{\bp}{\parm}h) \nn \\
&&\!\!\!\!\!\!\!\!\!\!\!\!-\fr{\parm}(\parm^2\bh\frac{\bp^2}{\parm^2}h)-\frac{\bp^2}{\parm^3}(h\parm^2\bh){\biggr ]}\ .
\eea
The last three lines of (\ref {result}) represent new quintic interaction vertices produced by the shift (\ref {shift}) acting on the cubic vertices in (\ref {lcg}). The expression for $A$ in (\ref {result}) contains terms of the form $h^3\bh^2$ thus representing the quintic MHV terms. Although $\bar A$ appears to be non-MHV it is not independent of $A$ since it is obtained by conjugation. Note that the entire first line of (\ref {finallag}) does not contribute to odd-point vertices thus making them free of time derivatives (and easier to extract) in this formalism.

\vskip 0.5cm
\section{Concluding Remarks}

The complete quintic interaction vertex in light-cone gravity has been derived. The result should prove useful in various ways as outlined in the introduction. It will be interesting to apply the field redefinitions of~\cite{AT} to this new interaction vertex. The hope is that these field redefintions, first introduced in~\cite{GR,PM}, will produce a MHV-like Lagrangian~\cite{CSW} for pure gravity to order $\kappa^3$. This should make the five-point KLT relations manifest~\cite{AT} at the level of the Einstein-Hilbert Lagrangian. In order to encounter ``truly" non-MHV structures, we need to move to six- and higher-point vertices. It should prove both non-trivial and instructive to derive the KLT relations, purely from field theory, for higher-point vertices where non-MHV structures abound.

\vskip 0.5cm
\subsubsection*{Acknowledgments}

I thank Stefano Kovacs, Hidehiko Shimada and Stefan Theisen for many valuable discussions and suggestions.

\end{document}